\documentclass{article}

\usepackage{amsmath}
\usepackage{amssymb}
\usepackage{graphicx}
\title{The Informative Herd: why humans and other animals imitate more when conditions are adverse}
\author{Alfonso P\'erez-Escudero\footnote{alfonso.perez.escudero@gmail.com - Cajal Institute (Consejo Superior de Investigaciones Cient\'ificas)} \ \& Gonzalo G. de Polavieja\footnote{gonzalo.polavieja@cajal.csic.es - Cajal Institute (Consejo Superior de Investigaciones Cient\'ificas)}}

\begin{document}

\maketitle

% Head 1
\begin{abstract}

Decisions in a group often result in imitation and aggregation, which are enhanced in panic, dangerous, stressful or negative situations. Current explanations of this enhancement are restricted to particular contexts, such as anti-predatory behavior \cite{Miller1922,Hamilton1971}, deflection of responsibility in humans \cite{Bradley1978,Zuckerman1979}, or cases in which the negative situation is associated with an increase in uncertainty \cite{Deutsch1955,Valone2002}. But this effect is observed across taxa and in very diverse conditions, suggesting that it may arise from a more general cause, such as a fundamental characteristic of social decision-making. Current decision-making theories do not explain it, but we noted that they concentrate on estimating which of the available options is the best one, implicitly neglecting the cases in which several options can be good at the same time \cite{Banerjee1992,Bikhchandani1992,Perez-Escudero2011}. We explore a more general model of decision-making that instead estimates the probability that each option is good, allowing several options to be good simultaneously \cite{Arganda2012}. This model predicts with great generality the enhanced imitation in negative situations. Fish and human behavioral data showing an increased imitation behavior in negative circumstances \cite{Hoare2004,Case2004} are well described by this type of decisions to choose a good option.
\end{abstract}

% Head 2
\section{Theory}

%The most widespread decision-making theories assume that subjects estimate which of the available options is the best one \cite{Banerjee1992,Bikhchandani1992,Perez-Escudero2011}. According to this view, even if all options are very similar, subjects will still make a cognitive effort to distinguish which one is the best. This assumption has the consequence that  subjects become insensitive to the overall quality of the options, only taking into account the differences between them. 

\begin{figure}
	\centering
		\includegraphics[width=1.00\columnwidth]{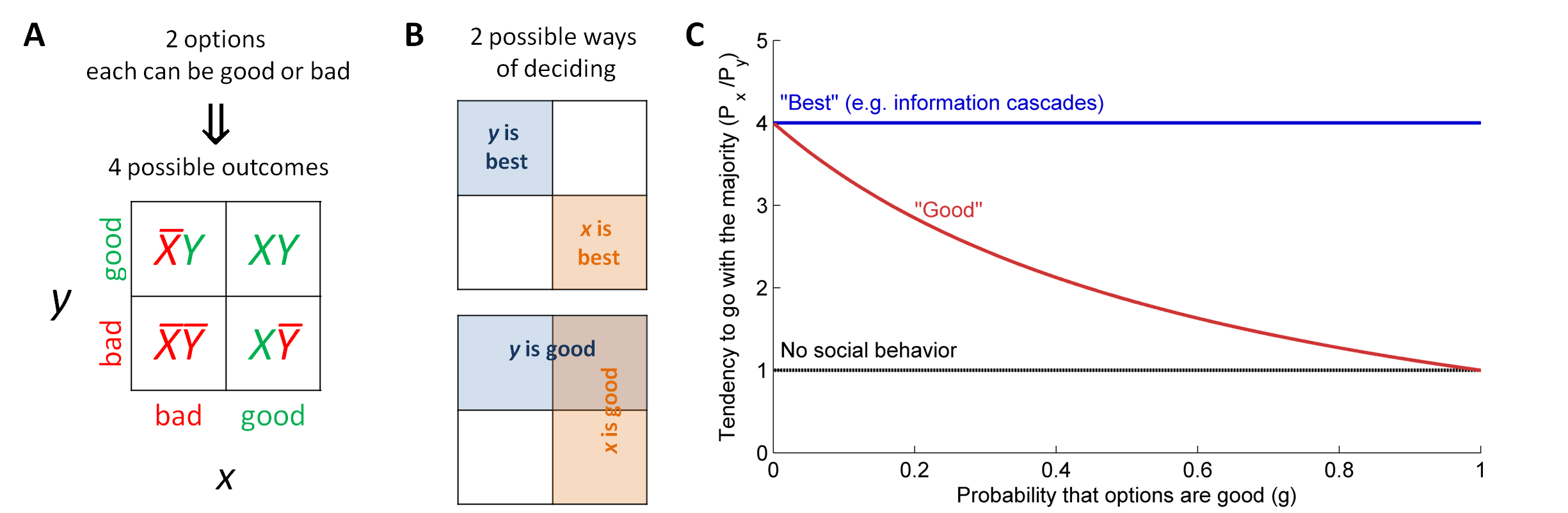}
	\caption{\textbf{Description of the theory.} \textbf{(A)} The subjects choose between two options, $x$ and $y$. Each option can be good or bad, so there are four possibilities (both options are good ($XY$), both options are bad ($\bar{X}\bar{Y}$), $x$ is good and $y$ is bad ($X\bar{Y}$) and $x$ bad, $y$ good ($\bar{X}Y$). \textbf{(B)} \textbf{Top:} When individuals estimate which option is best, they compare the probability that $x$ is good and $y$ is bad, $P(X\bar{Y})$, with the probability that $x$ is bad and $y$ is good, $P(\bar{X}Y)$. \textbf{Bottom:} When individuals estimate whether each option is good they compare the probability that $x$ is good, $P(XY)+P(X\bar{Y})$, with the probability that $y$ is good, $P(XY)+P(\bar{X}Y)$. \textbf{(C)} Tendency to follow the majority ($P_x/P_y$, assuming that the majority has chosen $x$ giving $P(B|X\bar{Y})/P(B|\bar{X}Y)=4$), as a function of the probability that the options are good (assuming that both options seem identical).}
	\label{fig:1}
\end{figure}

Let us consider the simple case of an individual choosing between two options, $x$ and $y$. Each option can be good or bad, so there are four possibilities: Both options are good ($XY$), both are bad ($\bar{X}\bar{Y}$), and one good and one bad ($\bar{X}Y$ and $X\bar{Y}$), Figure~\ref{fig:1}A.

The subject can use its private information ($C$) and the decisions made by other individuals ($B$) to estimate the probability for each of the four cases. For example, to estimate the probability that $x$ is good and $y$ is bad ($X\bar{Y}$) applying Bayes' theorem we get\footnote{Note that $P(B|X\bar{Y},C)=P(B|X\bar{Y})$ because $C$ represents information about the goodness of the options, so it does not add anything to the estimation when we restrict ourselves to $x$ being good and $y$ being bad, as in  $P(B|X\bar{Y},C)$.}
\begin{equation}\label{ecn:Bayes2}
P(X\bar{Y}|B,C)=\frac{P(B|X\bar{Y})P(X\bar{Y}|C)}{P(B|C)}.
\end{equation}

Private and social information are separated in the numerator of Equation~\ref{ecn:Bayes2}. The social term $P(B|X\bar{Y})$ measures how strongly the behaviors of the other individuals indicate that $x$ is good and $y$ is bad. We can expand this term to find an explicit expression \cite{Perez-Escudero2011,Arganda2012}, but it is not necessary here because our results will be independent of it. 

Private information is contained in the term $P(X\bar{Y}|C)$. Assuming that each option can be good or bad independently of the other we have $P(X\bar{Y}|C)= P(X|C)P(\bar{Y}|C)$. We consider the symmetric case, in which both options have equal probability of being good using only private information, $g\equiv P(X|C)=P(Y|C)$. Note that $P(\bar{Y}|C)=1-P(Y|C)=1-g$, so Equation~\ref{ecn:Bayes2} becomes
\begin{equation}\label{ecn:ProbXnoY}
P(X\bar{Y}|B,C)=\frac{P(B|X\bar{Y})g(1-g)}{P(B|C)}.
\end{equation}

An individual that wants to choose the best option needs to compare $P(X\bar{Y}|B,C)$ with $P(\bar{X}Y|B,C)$ (Figure~\ref{fig:1}B, top). The ratio between these two probabilities is
\begin{equation}\label{ecn:RatioBest}
\frac{P(x\text{ is the best})}{P(y\text{ is the best})}=\frac{P(B|X\bar{Y})g(1-g)}{P(B|\bar{X}Y)(1-g)g}=\frac{P(B|X\bar{Y})}{P(B|\bar{X}Y)}
\end{equation}
The result does not depend on the individual's private estimate of the quality of the options ($g$) (blue line in Figure~\ref{fig:1}C). Therefore, decision-making models based on finding the best choice fail to reproduce the increase of aggregation in adverse conditions. This is the case of Information Cascades \cite{Banerjee1992,Bikhchandani1992} and other similar models \cite{Perez-Escudero2011}.

An individual that wants to choose a good option needs to compare $P(x\text{ is good})=P(X\bar{Y}|B,C)+P(XY|B,C)$ with $P(y\text{ is good})=P(\bar{X}Y|B,C)+P(XY|B,C)$ (Figure~\ref{fig:1}B, bottom). The ratio between these two probabilities is
\begin{equation}\label{ecn:RatioGood}
\frac{P(x\text{ is good})}{P(y\text{ is good})}=\frac{P(B|X\bar{Y})g(1-g)+P(B|XY)g^2}{P(B|\bar{X}Y)(1-g)g+P(B|XY)g^2}.
\end{equation}

Equation~\ref{ecn:RatioGood} reproduces the experimental observations (Figure~\ref{fig:1}C, red line): When the individuals perceive that conditions are good ($g\rightarrow 1$ and the ratio tends to 1) they follow each other less than when they perceive that conditions are bad ($g\rightarrow 0$ and the ratio tends to the same value as Equation~\ref{ecn:RatioBest}).

\section{Tests of the theory}

The model accurately reproduces data of several species \cite{Arganda2012,Perez-Escudero2013}. Here we test it quantitatively in experiments that manipulate the overall quality of the options.

\begin{figure}
	\centering
		\includegraphics[width=1.00\columnwidth]{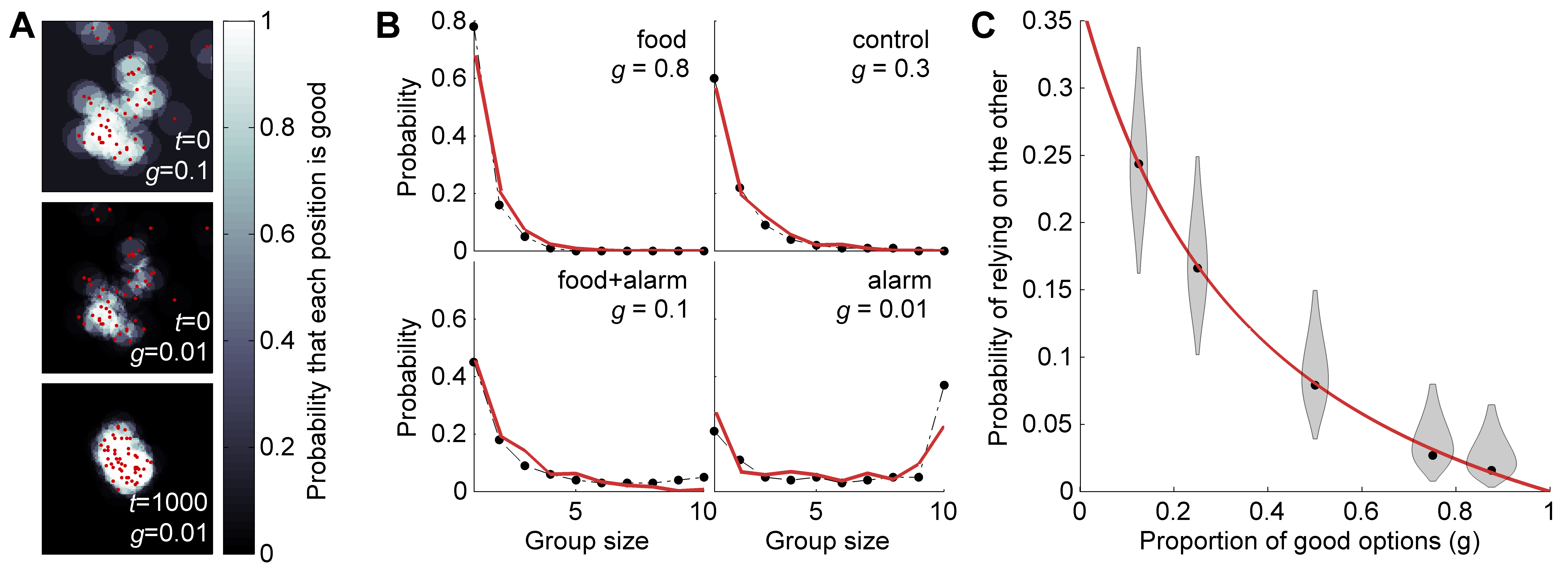}
	\caption{\textbf{Tests of the theory.} \textbf{(A)} Results of simulation for a control condition (top), just after a sudden change in private information (middle) and after some time (bottom). Red dots indicate the positions of the animals, and the color indicates the probability that each position is a good place according to the model. \textbf{(B)} Statistics of group sizes for the different conditions of the experiment by \cite{Hoare2004}. Experimental results (dots) and fit with the model (red lines, all parameters except $g$ remain constant in the four conditions). \textbf{(C)} Proability of relying on the opinion of another person as a function of the proportion of good cards, for the experimental results of \cite{Case2004} (dots) and the model (line).}
	\label{fig:2}
\end{figure}

\subsection{Group cohesion in fish}

To simulate the behavior of a group of free-moving animals we compute the probability that each position in space (each pixel in Figure~\ref{fig:2}A) is a good place. We use a model like Equation~\ref{ecn:RatioGood}, assuming that each animal contributes to the probability of all locations within a certain radius. Animals move at constant speed, choosing their destination with a probability given by the model \cite{Arganda2012}. In good conditions few individuals are enough to obtain a high probability that their surroundings are good (Figure~\ref{fig:2}A, top). If conditions suddenly deteriorate (for example because of an alarm signal), only high-density regions maintain a high probability of being good (Figure~\ref{fig:2}A, middle). Animals will tend to go to these regions, and after a short time the group becomes tighter (Figure~\ref{fig:2}A, bottom).
 
\cite{Hoare2004} studied the statistics of group sizes of killifish in conditions with different private information: control, food odour, alarm odour and a mixture food and alarm odours. They find that groups tend to be bigger when conditions are negative (Figure~\ref{fig:2}B, blue dots). We found a very good correspondence with this data by changing the parameter of private information in each condition, while keeping all other parameters constant across conditions (Figure~\ref{fig:2}B, red lines).

\subsection{Decision-making in humans}
A more stringent test of the theory can be done using experiments in which the value of the private information parameter $g$ is set experimentally, instead of being a parameter of the fit. \cite{Case2004} studied how much humans rely on social information depending on the probability of success. The subject must choose one card out of eight, getting a reward if it is a 'good' card. The subject knows the proportion of good cards, and she can either choose one card herself or rely on the opinion of another person. The results show the trend predicted by the model: Subjects rely on social information more when the probability that each option is good is low (Figure~\ref{fig:2}C, dots for experimental results, lines for the model in Equation~\ref{ecn:RatioGood} (see details in \cite{Arganda2012}).

%Note that in this experiment the probability that options are good is not directly related to stress. This highlights the generality of our theory, that is not related only to increases of stress or danger, but to any change in the probability that options are good.

\newpage

% Bibliography
\bibliographystyle{plain}
\bibliography{RefsInfoHerd_Arxiv}

\end{document}